\documentstyle[prd,aps,epsf,epsfig]{revtex} 
\begin{document}
\draft
\twocolumn[\hsize\textwidth\columnwidth\hsize\csname@twocolumnfalse\endcsname

\title{Heuristic derivation of continuum kinetic equations 
from microscopic dynamics}

\author
{{Kwan-tai Leung}\cite{email_ktl}}
\address{
Institute of Physics, Academia Sinica,
Taipei, Taiwan 11529, R.O.C.
}

\maketitle
\centerline{\small (Last revised \today)}

\begin{abstract}
We present an approximate and heuristic scheme for 
the derivation of continuum kinetic equations
from microscopic dynamics for stochastic, interacting systems.  
The method consists of a mean-field type, decoupled approximation 
of the master equation followed by the `naive' continuum limit.
The Ising model and driven diffusive systems are used as illustrations.
The equations derived are in agreement with other approaches,
and consequences of the microscopic dependences 
of coarse-grained parameters compare favorably with exact 
or high-temperature expansions.
The method is valuable when more systematic and rigorous
approaches fail, and when microscopic inputs in
the continuum theory are desirable.
\end{abstract}

\vspace{2pc}
]

\section{Introduction}\label{INTRO}

Ever since its introduction
in a classic treatment of the Brownian motion\cite{langevin},  
the Langevin equation has been playing an important 
role in modern statistical physics.
It provides a mathematical framework and a physical basis
for studying stochastic processes in statistical, mechanical systems.
Applications are wide-ranging\cite{vankampen}, including
chemical reactions, laser physics, diffusive processes,
and modern theories of dynamical critical phenomena\cite{hh}.
Recent topics such as surface growth\cite{surface}
and pattern formation\cite{pattern} 
also rely heavily on the Langevin equation.

It is fair to say, however, that despite its popularity, 
the Langevin equation for a specific problem 
is seldom derived from the corresponding microscopics.
It is often postulated on grounds of symmetry and physical reasoning.
Only rarely in simple circumstances is it derived,
for example, in reasonable details from the more
fundamental master equation.
In this article, we shall present an approximate 
scheme for deriving the Langevin equation, starting from 
the microscopic specification of the dynamics.
Our goal is not to provide a formal derivation, 
but rather to propose a simple and heuristic means that can be 
applied generally to many stochastic systems, 
several of which shall be discussed below.

Expositions of the historical, philosophical and technical 
aspects of the Langevin equation are beyond the scope of this article, 
interested readers are thus referred to the 
relevant literature\cite{vankampen}.
This paper is organized as follows:
An elementary recapitulation of the 
master equation and Langevin equation
is presented in Section \ref{TDGL}.
Section \ref{DERIVATION} contains several examples 
as illustrations of the method, as well as assessment
of the quality of the approximation involved.
Conclusion is given in Section \ref{CONCLUSION}.
A discussion of the noise correlation is given in the Appendix.

\section{master equation and time-dependent 
Ginzburg-Landau equation}\label{TDGL}

In statistical physics, one of the most important applications 
of the Langevin equation is 
in the theories of dynamical critical phenomena\cite{hh}.
Therefore, our discussion shall be cast in that language,
although it should be obvious that the method itself is not limited
to systems exhibiting those phenomena.
For concreteness, consider the kinetic Ising model 
that obeys Glauber (i.e., spin-flip) dynamics\cite{Glauber}.  
At the classical, microscopic level of description,
the system consists of $N$ spins $\sigma_i$ 
interacting {\em via\/} the Hamiltonian
\begin{equation}
H=-J \sum_{\langle i,j \rangle} \sigma_i \sigma_j,
\label{ham}
\end{equation}
where $J$ is the coupling constant, and 
$\langle i,j \rangle$ denotes a sum over nearest-neighbor pairs.
The time evolution of the system is governed by a master equation
\begin{eqnarray}
&& P({\vec\sigma};t+1)-P({\vec\sigma};t)= 
\nonumber\\ &&\qquad 
\sum_{\vec\sigma'}
         \left[ w({\vec\sigma'}\to {\vec\sigma})P({\vec\sigma'};t)
              -w({\vec\sigma}\to {\vec\sigma'})P({\vec\sigma};t) \right],
\label{master}
\end{eqnarray}
where $P({\vec\sigma};t)$ is the joint probability of finding the
system in the spin configuration 
$\vec\sigma\equiv\{\sigma_1,\sigma_2,\cdots,\sigma_N\}$
at time $t$, and $w$'s are the transition rates
between two configurations that differ only by one spin flip.
There is a great deal of freedom in the choice of $w$,
as long as the following detailed balance condition 
is satisfied to ensure the same equilibrium distribution 
$P_{\rm eq}({\vec\sigma})\sim e^{-\beta H(\vec\sigma)}$
\begin{equation}
 {w({\vec\sigma'}\to {\vec\sigma}) \over
  w({\vec\sigma}\to {\vec\sigma'})      }
={P_{\rm eq}({\vec\sigma}) \over P_{\rm eq}({\vec\sigma'}) }
=e^{-\beta[H(\vec\sigma)-H(\vec\sigma')]},
\label{db}
\end{equation}
with $\beta=1/k_B T$.
In practice, the choice is largely dictated by mathematical convenience.
The most common choices of $W$ are the Metropolis rate in Monte Carlo
simulations for its ease of implementation, 
and the heat-bath rate (also known as the Kawasaki
rate\cite{kawasaki}) in analytic calculations for its analyticity.
In this paper, we confine our attention to the latter choice.
It is given by
\begin{equation}
w({\vec\sigma}\to {\vec\sigma'})=
       { 1 \over 1+e^{-\beta[H(\vec\sigma)-H(\vec\sigma')]} }.
\label{heatbath}
\end{equation}

Since the master equation is not very convenient for 
analytic purposes such as a renormalization group analysis, 
one often turns to a mesoscopic, continuum representation.
For an Ising system with Glauber dynamics, 
the relevant continuum field is 
the local magnetization density $\phi(\vec{r},t)$, which
obeys a kinetic equation
\begin{eqnarray}
{\partial\phi\over \partial t}
&=& -\Gamma {\delta {\cal H}\over \delta \phi}+\zeta,
\label{tdgl}\\
{\cal H} &=&
\int d^d r\,\left\{ {1\over 2}(\nabla\phi)^2 + V(\phi) \right\},
\label{hamcg}\\
V(\phi) &=&
 {u\over 2}\phi^2 + {g\over 4!}\phi^4 + \cdots.
\label{Vcg}
\end{eqnarray}
This is an example of the time-dependent Ginzburg-Landau (TDGL) 
kinetic equation.
In (\ref{hamcg}), $d$ is the dimensionality of the system,
and $\cal H$ is a coarse-grained Hamiltonian. 
For (\ref{tdgl}) to describe a stochastic process, the noise term
$\zeta(\vec{r},t)$ is needed, which 
accounts for the effect of thermal fluctuations and prevents
the system from trapping in metastable states.
For mathematical convenience, it is often taken to be Gaussian 
with zero mean:
\begin{eqnarray}
\langle \zeta(\vec{r},t) \rangle&=&0, \\
\langle \zeta(\vec{r},t)\zeta(\vec{r}',t') \rangle&=&
2{D} \delta(\vec{r}-\vec{r}')\delta(t-t').
\label{noisecorrel}
\end{eqnarray}
For equilibrium systems, 
the correlation $D$ in (\ref{noisecorrel}) 
has to be chosen to ensure that
the stationary solution of Eq.~(\ref{tdgl}) is consistent with
the Boltzmann weight ${\cal P}_{\rm eq}\sim e^{-{\cal H}}$ (cf. Appendix).
For a system as simple as the Ising model,
the static continuum Hamiltonian $\cal H$ 
can actually be derived from the microscopic $H$ 
{\em via\/} the partition function
by means of the Hubbard-Stratonovich transformation\cite{amit},
which is a trick based on the Gaussian integral.
For the dynamics, the TDGL equation can be derived by 
coarse graining the master equation\cite{langer71,omegaexp}, 
which in principle yields expressions of the coarse-grained
parameters $\Gamma$, $u$ and $g$ in (\ref{tdgl})-(\ref{Vcg}) 
as functions of microscopic ones in (\ref{master}).

However, for more complicated $H$,
the Hubbard-Stratonovich transformation fails and
the coarse graining cannot be done explicitly.
Moreover, these methods rely on the existence of a Hamiltonian
$H$ and the associated equilibrium Boltzmann weight $e^{-\beta H}$,
which is not valid in generic non-equilibrium situations
defined only by the dynamics\cite{ness}.
For these reasons, it is highly desirable to
have a way with which a continuum description can be
extracted directly from the dynamics\cite{intuitive}.

\section{factorization and naive continuum expansion}\label{DERIVATION}

Our method is very simple. It consists of two steps:
a mean-field type factorization of joint probabilities 
into singlet ones in the master equation, 
followed by a `naive' continuum expansion. 
The result is a continuum kinetic equation
with full knowledge of the microscopic dependence of 
the coarse-grained parameters.  Since the input is the master equation, 
whether the system is an equilibrium one or not\cite{ness} 
is irrelevant.  To illustrate, we now discuss several examples
in increasing order of sophistication.

\subsection{1D Ising model}

Focusing on a spin at position $x$
in a one-dimensional (1D) Ising model, it is easy to find by 
integrating out all other spins in Eq.~(\ref{master}) that
\begin{eqnarray}
 P_+&&(x;t+1)-P_+(x;t)= 
     P_{---}w_{---}
    +P_{--+}w_{--+}
\nonumber\\
&&
    +P_{+-+}w_{+-+}
    +P_{+--}w_{+--}
    -P_{-+-}w_{-+-}
\nonumber\\
&&
    -P_{-++}w_{-++}
    -P_{++-}w_{++-}
    -P_{+++}w_{+++},
\end{eqnarray}
where $P_+(x;t)$ denotes the singlet probability of finding the spin up
at site $x$ at time $t$, and
$P_{+++}(x;t)$ denotes the joint probability of finding three spins up
at site $x-1$, $x$, $x+1$ respectively, and so on.
From (\ref{heatbath}), the heat-bath transition rates are given by
$w_{---}=w_{+++}=W_4$,
$w_{--+}=w_{+--}=w_{-++}=w_{++-}=W_0$,
$w_{+-+}=w_{-+-}=W_{-4}$, where 
\begin{equation}
W_n\equiv { 1 \over 1+e^{n\beta J} }.
\label{w_n}
\end{equation}
Adopting a mean-field approximation, the joint probabilities
are replaced by their factorizations, e.g.,
$P_{++-}(x;t)\to P_+(x-1;t)P_+(x;t)P_-(x+1;t)$.
Since $\sum_\sigma \sigma P_\sigma(x;t)=P_+(x;t)-P_-(x;t)
=\langle \sigma\rangle$,
and $\sum_\sigma P_\sigma(x;t)=P_+(x;t)+P_-(x;t)=1$,
in the spirit of coarse graining 
we proceed to make the identification 
\begin{equation}
P_\pm(x;t) \leftrightarrow {1\pm \phi(x;t)\over 2},
\label{P2phi}
\end{equation}
where $\phi$ is the local magnetization density.
By using $\phi$ instead of spin number densities, we take
advantage of symmetries anticipated in the final kinetic equation.
Since a spin flip depends on a total of $z+1$ spins in (\ref{master}),
where $z=2d$ for hyper-cubic lattices,
the factorization effectively produces a power series expansion
in $\phi$ up to $\phi^{z+1}$.
After replacing $P$'s by $\phi$'s,  
we make the transition to the continuum by `naively'
expanding about $x$, such as:
\begin{equation}
\phi(x\pm 1;t)\rightarrow \phi(x;t)
              \pm {\partial \phi(x;t)\over \partial x}
              +{1\over2}{\partial^2 \phi(x;t)\over \partial x^2} +\cdots.
\label{naive}
\end{equation}
For most applications, we are only interested in the
long-distance behavior, hence it suffices to stop at the lowest
derivatives as shown.
This procedure results in 
a deterministic kinetic equation for $\phi$ in precisely the form of 
(\ref{tdgl})-(\ref{Vcg}), barring the noise term $\zeta$:
\[
{\partial \phi\over \partial t}
=-\Gamma\left(-{\partial^2\phi\over\partial x^2}+r\phi
              +{ g\over 6}\phi^3\right),
\]
where the coefficients are given by
\begin{eqnarray}
\Gamma&=&{1\over 2} (W_{-4} - W_4),
\nonumber\\
r&=&{ 1\over 2\Gamma}\left( 3W_4-W_{-4}+2W_0 \right),
\nonumber\\
g&=&{ 3\over \Gamma} \left( W_4+W_{-4}-2W_0 \right).
\label{1DIsing}
\end{eqnarray}

Several remarks are in order:
\begin{enumerate}
\item
Symmetries in the resulting 
continuum equation (with respect to $\phi$, $x$ and $t$)
are as expected, because the approximations respect those symmetries
and leave them intact.
\item
There are explicit temperature dependences in the coefficients 
which cannot be deduced by symmetry or physical reasoning.
Such dependences are specific to the choice of jump rates
which manifests through the approximations used.
\item
Noting that $W_{-n}=1-W_n$ for any $n$,  we find
$\Gamma={1\over 2}-W_4>0$ and $g=0$ at any $T$\cite{truncate}, and
$r=2 W_4/\Gamma$ has one zero, at $T=0$.
This is consistent with the absence of 
phase transition in the 1D Ising model at any finite temperature, 
an improvement over the usual mean-field result 
$T_c^{\rm MF}=2 J/k_B$.
There is no stability problem arising from $g=0$
because the quadratic coefficient is positive for $T>0$.
\item
In the presence of an external magnetic field $h$, the degeneracies
in jump rates are lifted 
(e.g., $W_{+\pm-}=(1+e^{\pm 2\beta h})^{-1}$). 
To $O(h)$, the kinetic equation acquires a new term 
$\Gamma\mu h$ on the right hand side, where 
\begin{equation}
\mu={2\beta \over \Gamma} (W_0^2 + W_4 W_{-4}).
\end{equation}
Linear response then determines that the susceptibility
is $\chi=\mu/r=\beta (1-\gamma^2/2)/(1-\gamma)$, 
where $\gamma\equiv \tanh 2\beta J$. 
In the Appendix we show that $\mu$ is 
needed to fix the noise correlation.
\item
Besides capturing the correct symmetries,
our results compare quite well 
with exact results.  From (\ref{tdgl}), the
relaxation time can be read off easily as
$\tau=1/\Gamma r=1/(1-\gamma)$.
This turns out to be exact\cite{Glauber}.
For the susceptibility, deviation from
the exact result $\beta e^{2\beta J}$\cite{Glauber} 
shows up only at $O((\beta J)^5)$ when expanded in $\beta J$.
Hence, our method has the advantage that 
it embodies a refined mean-field theory, 
as already applied to studies of stochastic resonance 
in Ising systems\cite{leungneda}.
\item
Finally, due to the factorizations only the deterministic terms
in (\ref{tdgl}) can be derived. The noise term has 
to be deduced separately (see Appendix).
The result for the noise correlation $D$ in (\ref{noisecorrel})
is ${D}=k_B T \mu \Gamma$.
\end{enumerate}

Having gone through the details of our method, we now turn 
to a few less trivial examples.

\subsection{2D Ising model}

The same procedure can be applied to the 2D Ising model 
with Glauber dynamics.
Again we obtain (\ref{tdgl}), with the parameters given by
\begin{eqnarray}
\Gamma&=&{1\over 8}  (      -2W_4 + 2W_{-4} -  W_8 + W_{-8}),
\label{Tdep_G2d}\\
r&=&{1\over 8 \Gamma}( 6W_0+12W_4 - 4W_{-4} + 5W_8 -3W_{-8}),
\label{Tdep_u2d}\\
g&=&{3\over 2 \Gamma}(-6W_0-4 W_4 + 4W_{-4} + 5W_8 + W_{-8}),
\label{Tdep_g2d}\\
\mu&=&{\beta \over 2\Gamma} (3W_0^2+4W_4 W_{-4} + W_8 W_{-8}).
\label{Tdep_mu2d}
\end{eqnarray}
It is worth noting that $\Gamma$, $g$ and $\mu$ 
are positive definite for all $T>0$, whereas $r$ has one zero at
$T_c^{\rm GL} \approx 3.0898 J/k_B \approx 1.3616 T_c$, 
again an improvement over the mean-field prediction 
$T_c^{\rm MF}=4 J/k_B$, 
where $T_c=-2J/k_B\ln(\sqrt{2}-1)\approx2.2692J/k_B$ 
is the exact critical temperature.
As expected, there is no $\phi^5$ and higher order term\cite{truncate}.

The results of $\tau$ and $\chi$ for the Gaussian case
($g=0$) are quite satisfactory.  They differ from high-temperature
series expansions\cite{hte} at order $O((\beta J)^5)$ and
$O((\beta J)^4)$, respectively,  whereas the usual mean-field
results are worse, at $O((\beta J)^3)$ and $O((\beta J)^2)$.

\subsection{3D Ising model}

Despite being more tedious (128 terms on the right-hand side 
of the master equation),
we also derive the kinetic equation 
for the 3D Ising model with Glauber dynamics. The results are:
\begin{eqnarray}
\Gamma&=&{1\over 32}  
         (W_{-12} +4W_{-8}+5W_{-4}-5W_4 -
\nonumber \\ &&\qquad 
          4 W_8- W_{12}),
\label{Tdep_G3d}\\
r&=&{1\over 32 \Gamma}  
         (-5W_{-12} -18W_{-8}-15W_{-4}+20W_{0}+
\nonumber \\ &&\qquad 
          45W_4+30W_8+7W_{12}),
\label{Tdep_u3d}\\
g&=&{15\over 16\Gamma}
         (-W_{-12} +6W_{-8}+9W_{-4}-12W_{0}-
\nonumber \\ &&\qquad 
         15W_4+6W_8+7W_{12}),
\label{Tdep_g3d}\\
\mu&=&{\beta \over 8\Gamma} 
         (10W_{0}^2+15W_{-4}W_4+6W_{-8}W_8+
\nonumber \\ &&\qquad 
          W_{-12}W_{12}),
\label{Tdep_mu3d}
\end{eqnarray}
As for 2D Ising,  $\Gamma$, $g$ and $\mu$ 
are positive definite for all $T>0$, and $r$ has one zero at
$T_c^{\rm GL} \approx 5.0733 J/k_B $, 12\% higher than 
the best estimate\cite{landau},
compared to $T_c^{\rm MF}=6 J/k_B$. 

\subsection{1D driven lattice gas}

In many generic non-equilibrium systems\cite{ness},
the free energy does not exist and one has to start from
the dynamics, such as described by the master equation.
A notable example is the driven diffusive system\cite{dds}, which
is regarded as a paradigm of spatially extensive 
interacting systems that exhibit cooperative phenomena 
in steady-state non-equilibrium situations.
In its standard form, it models an Ising-like lattice gas of particles
whose motion along a certain direction 
is biased by an external drive denoted by $E$.
For $E=0$, the model reduces to the ordinary kinetic Ising model
with Kawasaki, or spin-exchange, dynamics (model B in \cite{hh}).

A question subject to recent debate concerns the form of 
nonlinearities associated with $E$\cite{debate1,debate2}.
That is an important issue because the nonlinearities decide 
to which universality class of critical behavior the system belongs.
It is interesting to see what the present method says about that.
First, we consider a one-dimensional, simplified version in which
the particles are not interacting except being hard-core, 
but their hoppings to nearest neighbors are biased by having
different jump rates, $p$ and $q$, to the right and left respectively.
Hence, the master equation reads
\begin{eqnarray}
&& P_+(x;t+1)-P_+(x;t)=
\nonumber\\ &&\qquad  
p P_{+-}(x-1;t) + q P_{-+}(x;t)
\nonumber\\ &&\qquad  
-p P_{+-}(x;t) -q P_{-+}(x-1;t),
\label{1Dddsmaster}
\end{eqnarray}
where as usual an up(down) spin corresponds to the occupation 
of a particle(hole), and joint probabilities such as 
$P_{+-}(x-1;t)$ means the probability
of finding a particle-hole pair at site $x-1$ and $x$.
After factorizations and applications of
(\ref{P2phi}) and (\ref{naive}), we readily find
\begin{equation}
{\partial \phi \over \partial t}=
{\cal D} {\partial^2 \phi \over \partial x^2} 
+ {{\cal E}\over 2} {\partial \phi^2 \over \partial x}
\label{1Dddstdgl}
\end{equation}
where the diffusion coefficient is ${\cal D}=(p+q)/2$, as expected,
and the coefficient of driving is ${\cal E}=(p-q)$.
The nonlinear term is the same as 
in the `standard' field theoretic model\cite{ddsft} which was
proposed on grounds of symmetries.
As side remarks, note that we obtain the diffusion equation for $p=q$, 
and that $\cal E$ is smooth in the 
`infinite' drive limit ($p=1$, $q=0$) which is used in most
Monte Carlo simulations of driven diffusive systems.

\subsection{2D driven lattice gas}

Generalization of the previous result to
the 2D interacting driven lattice gas is immediate, 
despite the unpleasant fact that there are altogether 
512 terms in the master equation.
In the presence of a drive $E$ along the $+y$ direction and 
attractive ($J>0$ in (\ref{ham})) interaction between particles,
the heat-bath rates for hoppings of particles
along and against the drive take the form
\begin{equation}
W_{n,\pm E}\equiv { 1 \over 1+e^{n\beta J \mp E \beta J} },
\label{w_ne}
\end{equation}
where the dimensionless $E$ ($0\leq E < \infty$)
represents the ``work done'' on the particle by the field. 
Obviously, the rates for hoppings perpendicular to $E$ are $W_{n,0}=W_n$.

\begin{figure}[htp]
\epsfig{figure=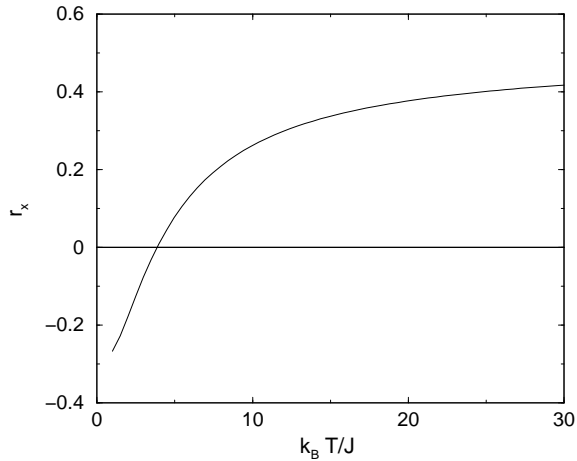,width=3.0in,angle=-0}
\caption{
Quadratic coefficient in the transverse direction $r_x$ 
plotted vs. temperature. 
Its zero locates the critical temperature $T_c^{\rm GL}$.
}
\label{fig:rx}
\end{figure}

\begin{figure}[htp]
\epsfig{figure=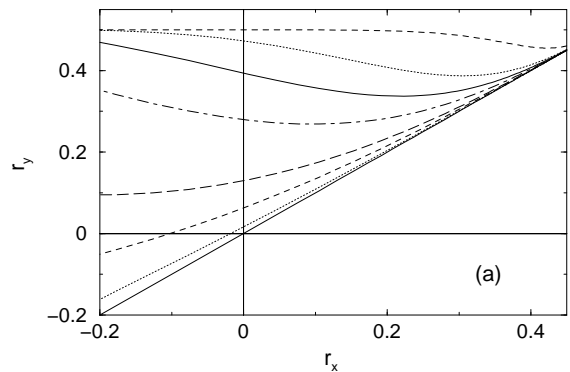,width=3.0in,angle=-0}
\epsfig{figure=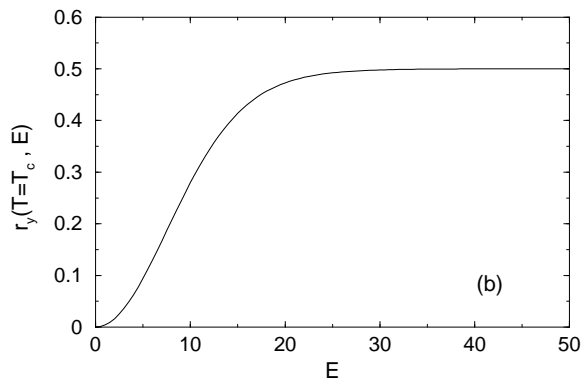,width=3.0in,angle=-0}
\caption{
(a) Trends of $r_y$ vs. $r_x$ as $T$ is varied across $T_c^{\rm GL}$
at fixed $E$. From bottom to top: $E=0$, 2, 4, 6, 10, 14, 20, and 50.
(b) Intercept $r_y(r_x=0)$ in (a) plotted vs. $E$.
}
\label{fig:ry}
\end{figure}

\begin{figure}[htp]
\epsfig{figure=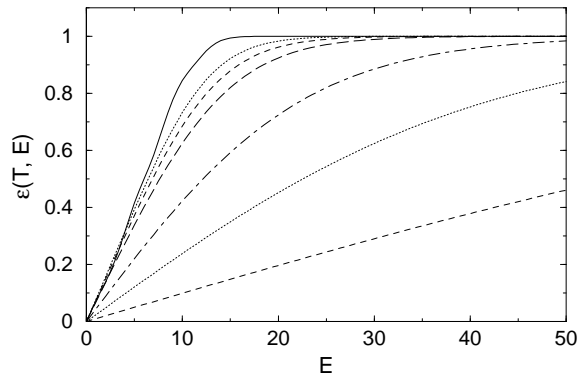,width=3.0in,angle=-0}
\caption{
The coefficient of the leading nonlinearity, $\cal E$ vs. 
the microscopic drive $E$ at
different temperatures. From top to bottom:
$k_B T/J=1$, 3, $3.86143(=T_c^{\rm GL}$), 5, 10, 20 and 50. 
}
\label{fig:E}
\end{figure}

Going through the same procedure as above,
we eventually obtain a kinetic equation which is in complete
agreement with the standard field theory of 
the driven diffusive system\cite{ddsft}: 
\begin{eqnarray}
{\partial \phi\over \partial t}
&=& 
-\left( \alpha_x    {\partial^4\over \partial x^4}
       +\alpha_{xy} {\partial^4\over \partial x^2\partial y^2}
       +\alpha_{y}  {\partial^4\over \partial y^4} \right)\phi
\nonumber \\ && 
+\left( r_x {\partial^2\over \partial x^2}
       +r_y {\partial^2\over \partial y^2}\right)\phi
+{1\over 6} \left( g_x{\partial^2\over \partial x^2}
                  +g_y{\partial^2\over \partial y^2}\right) \phi^3
\nonumber \\ &&
+{{\cal E}\over 2} {\partial \phi^2 \over \partial y}.
\label{dds2d}
\end{eqnarray}
The anisotropies are generated by the drive.
Excluding the last term, this is the anisotropic generalization
of the deterministic TDGL equation 
with conserved magnetization, i.e., model B\cite{hh}:
\begin{equation}
{\partial \phi\over \partial t} = 
\nabla^2\left( -\alpha \nabla^2\phi + r \phi +{g\over 6} \phi^3\right).
\label{modelB}
\end{equation}
All coefficients are determined:
\begin{eqnarray}
\alpha_x &=& 
{1\over 384}(69 - 85 W_4 - 68 W_8 - 17 W_{12}),
\\
\alpha_{xy} &=& 
{1\over 256}
(20 - 20\,W_4 - 16\,W_8 - 4\,W_{12} + W_{-12,-E} + 
\nonumber \\ &&
W_{-12,E} + 4\,W_{-8,-E} + 4\,W_{-8,E} + 
\nonumber \\ &&
5\,W_{-4,-E} + 5\,W_{-4,E} - 5\,W_{4,-E} - 5\,W_{4,E} - 
\nonumber \\ &&
4\,W_{8,-E} - 4\,W_{8,E} - W_{12,-E} - W_{12,E}),
\label{dds2d_alpxy}
\\
\alpha_{y} &=& 
{1\over 768}
(8\,W_{-12,-E} + 8\,W_{-12,E} + 31\,W_{-8,-E} + 
\nonumber \\ &&
31\,W_{-8,E} + 35\,W_{-4,-E} + 35\,W_{-4,E} - 
\nonumber \\ &&
10\,W_{0,-E} - 10\,W_{0,E} - 50\,W_{4,-E} - 50\,W_{4,E} - 
\nonumber \\ &&
37\,W_{8,-E} - 37\,W_{8,E} - 9\,W_{12,-E} - 9\,W_{12,E}),
\label{dds2d_alpy}
\\
r_x &=& 
{1\over 32}(-9 + 25\,W_4 + 20\,W_8 + 5\,W_{12}),
\label{dds2d_rx}
\\
r_y &=& 
{1\over 64}
(-2\,W_{-12,-E} - 2\,W_{-12,E} - 7\,W_{-8,-E} - 
\nonumber \\ &&
7\,W_{-8,E} - 5\,W_{-4,-E} - 5\,W_{-4,E} + 10\,W_{0,-E} + 
\nonumber \\ &&
10\,W_{0,E} + 20\,W_{4,-E} + 20\,W_{4,E} + 13\,W_{8,-E} + 
\nonumber \\ &&
13\,W_{8,E} + 3\,W_{12,-E} + 3\,W_{12,E}),
\label{dds2d_ry}
\\
g_x &=& 
{5\over 16} ( 3 + W_4 - 4\,W_8 - 3\,W_{12} ),
\label{dds2d_gx}
\\
g_y &=& 
{1\over 32}
(6\,W_{-12,-E} + 6\,W_{-12,E} + 7\,W_{-8,-E} + 
\nonumber \\ &&
7\,W_{-8,E} - W_{-4,-E} - W_{-4,E} + 6\,W_{0,-E} + 
\nonumber \\ &&
6\,W_{0,E} + 4\,W_{4,-E} + 4\,W_{4,E} - 13\,W_{8,-E} - 
\nonumber \\ &&
13\,W_{8,E} - 9\,W_{12,-E} - 9\,W_{12,E})
\label{dds2d_gy}
\\
{\cal E}&=& 
{1\over 16}
(-W_{-12,-E} + W_{-12,E} - 3\,W_{-8,-E} + 
\nonumber \\ &&
3\,W_{-8,E} - 3\,W_{-4,-E} + 3\,W_{-4,E} - 
\nonumber \\ &&
2\,W_{0,-E} + 2\,W_{0,E} - 3\,W_{4,-E} + 3\,W_{4,E} - 
\nonumber \\ &&
3\,W_{8,-E} + 3\,W_{8,E} - W_{12,-E} + W_{12,E}).
\label{dds2d_e}
\end{eqnarray}
They have these important properties:
\begin{enumerate}
\item
All but $\cal E$ are even in $E$,  consistent with the
invariance of the dynamics under $\{E\to -E, y\to -y\}$.
\item
The quadratic coefficient $r_x$ is independent of $E$. 
It has one zero at $T_c^{\rm GL}=3.86143 J/k_B$, 
as shown in Fig.~\ref{fig:rx}.
In contrast, $r_y$ depends on $E$. 
Fig.~\ref{fig:ry} displays the behavior 
of $r_y(T,E)$ versus $r_x(T)$ as $T$ is
lowered from above $T_c^{\rm GL}$ at fixed $E$,
as well as $r_y(T=T_c^{\rm GL},E)$ versus $E$.
It shows that for any $E>0$, 
$r_x$ always vanishes before $r_y$ does when $T$ is decreased.
For small $E$, $r_y\approx r_x + c E^2$ where $c>0$.
Consequently, at the critical temperature, the dominant
derivatives come from the $r_y$ and $\alpha_x$ terms,
leading to the identification of an intrinsically anisotropic 
critical theory with scaling of momenta $k_y\sim k_x^2$. 
This agrees with a previous perturbative argument\cite{ddsft}. 
\item
The coefficient $\cal E$ of the leading nonlinearity induced
by the drive vanishes linearly in $E$ at small $E$, and saturates
to a constant at $E=\infty$. This dependence, exhibited in 
Fig.~\ref{fig:E}, is already anticipated above from the 1D model 
and argued previously\cite{debate2},
but at odds with the claims in \cite{debate1}.
\item
At $E=0$: we find $\alpha_x=\alpha_y \neq \alpha_{xy}$--
the continuum model derived is not rotationally invariant.
This is not surprising because the initial lattice model is not.
However, it turns out that $\alpha_{xy}=\alpha_x$ at $T_c^{\rm GL}$.
At present, we are not sure whether this acquirement
of higher symmetry at the critical point is general for
the underlying lattice model or specific to the method.
\end{enumerate}

\subsection{two-species driven lattice gas}

In all the above examples, each site has only two local states
(spin up or down).
In the two-species driven lattice gas model\cite{shz}, 
motivated in part by multi-ionic conductors and traffic flow problems,
there are three possibilities: 
as a hole, or either of two types of particles.
The two types of particles are driven in opposite directions
as if they were oppositely charged and driven by an electric field,
with local particle densities denoted by $\rho_+$ and $\rho_-$.
Due to the extra local state, it is not easy to write
down the correct set of equations by symmetry and intuition alone.
One way to proceed\cite{shz} 
is to express the entropy in terms of 
$\rho_+$ and $\rho_-$, and obtain the diffusion terms by 
functional differentiations. 
The driving terms can then be added to the kinetic equations
by generalizing that for the one-species model.

Another way is to apply the current method\cite{leungzia97}.
The equations derived are the same as in \cite{shz},
with the advantage of tractable microscopic origins 
in each coefficients.
Hence, this model further testifies the usefulness of the present 
approach when symmetry and intuition are not very helpful.

We end this section with a final remark.
Since the method begins with factorization of joint probabilities, 
the ensuing equation is deterministic, all information about
correlations seem to have lost. 
The situation can be remedied, however, by introducing a noise term 
to restore a probabilistic description. 
Correlations can then be computed by averaging over the noise
by means of standard field-theoretic techniques\cite{msr}.
For equilibrium systems, the noise can be fixed by 
requirements such as the fluctuation-dissipation theorem.
For nonequilibrium systems, there is no general rule. 
One usually has to extrapolate by analogy to equilibrium.

\section{Conclusion}\label{CONCLUSION}

We have presented in details a very simple and straight forward
method to derive the deterministic 
kinetic equations from known microscopic dynamics 
for stochastic, interacting systems.
The method has a mean-field flavor. 
It preserves the underlying symmetries of the dynamics
and is in line with the spirit of coarse graining.
The resulting equations are in good agreement with other
either more or less rigorous approaches,
as demonstrated explicitly {\em via\/} several examples.
Hence, despite the approximate and heuristic nature of our approach, 
it proves to be a useful and convenient means to obtain
a correct continuum theory, especially 
(i) when other more rigorous approaches do not apply;
(ii) when symmetries of the system is not intuitively obvious; and
(iii) when microscopic dependences of the continuum parameters are wanted.

\vspace{0.5cm}
\noindent{\bf{Acknowledgments}:}
This work is supported by the National Science Council of R.O.C.
under grant number NSC89-2112-M-001-015,
and a Main-Theme Grant of the Academia Sinica.

\appendix
\section{noise correlation}

Since the present method only gives
the deterministic part of the kinetic equation,
the noise term has to be considered separately.  
Here we follow the common practice to assume that 
the noise $\zeta$ is Gaussianly distributed 
and correlated over negligible ranges.
Then the only question is to determine
its correlation $D$ in (\ref{noisecorrel}).

There are two ways to do it. The first makes use of the 
correspondence between the Langevin equation 
\begin{eqnarray}
{\partial \phi\over \partial t}&=&
-\Gamma {\delta {\cal H}\over \delta \phi} + \zeta
\label{app:tdgl} \\
\langle \zeta(\vec{r},t)\zeta(\vec{r}',t') \rangle&=&
2{D} \delta(\vec{r}-\vec{r}')\delta(t-t'),
\label{app:noise}
\end{eqnarray}
and the Fokker-Planck equation
\begin{equation}
{\partial {\cal P}\over \partial t}=
-\int \, d^dx {\delta \over \delta \phi} 
\left( -\Gamma {\delta {\cal H} \over \delta \phi} {\cal P} 
-{D} {\delta {\cal P}\over \delta \phi} \right),
\label{app:FP}
\end{equation}
which is a continuity equation.
A stationary solution of the Fokker-Planck equation is obtained by
setting zero the probability current, i.e., $(\cdots )=0$, which gives
${\cal P} \propto e^{-\Gamma {\cal H}/{D}}$.
Since the free energy 
${\cal F}[h]$ in the presence of an external field $h$
is of the form ${\cal F}[h]={\cal F}[0] - h \phi$,
it differs from $\cal H$ by a factor of $\mu$.
Hence, by matching $e^{-{\cal F}[h]/k_BT}$
and $e^{-\Gamma {\cal H}[h]/{D}}$,
we deduce that
\begin{equation}
{D}=k_B T\mu\Gamma.
\label{app:N}
\end{equation}
In passing, it is worth noting that 
the kinetic coefficient defined in 
\begin{equation}
{\partial \phi\over \partial t}=
-\lambda {\delta {\cal F}\over \delta \phi} + \zeta
\label{app:tdgl2}
\end{equation}
is $\lambda={D}/k_B T$: the Einstein relation.

An alternative way to determine $D$ 
is to use the fluctuation-dissipation theorem, 
which in momentum-frequency space takes the form 
\begin{equation}
{ 2 k_B T\over \omega} {\rm Im} \,\chi (k,\omega) = G (k,\omega).
\label{app:FDT}
\end{equation}
Although neither the susceptibility $\chi$ nor the two-point correlation
function $G$ can be calculated in closed form for general
$\cal H$, (\ref{app:FDT}) holds order by order so that we only
need to consider the Gaussian model in the case of $g=0$.
Thus, by Fourier transforms, we obtain
$\chi(k,\omega)=\Gamma \mu/[-i\omega + \Gamma (k^2+r)]$
and $G(k,\omega)=2{D}/[\omega^2 + \Gamma^2 (k^2+r)^2]$,
which by virtue of (\ref{app:FDT}) also gives (\ref{app:N}).




\begin{references}
\bibitem
{omegaexp}
The Langevin equation can also be obtained by correspondence
with the Fokker-Planck equation which
can be derived from an expansion of the master equation 
in inverse powers of the system volume.  
However,  we are not aware of such a self-contained and explicit
derivation which gives the Ginzburg-Landau form without making
any assumption.
\bibitem
{ness}
By `generic' we refer to systems which have nonzero probability current
in configuration space (the terms enclosed by brackets in (\ref{app:FP})) 
even in stationary states. In the master equation,
this corresponds to the breakdown of detailed balance, i.e., 
the right-hand side of (\ref{master}) does not vanish 
term-by-term in stationary states but only after summation.
Usually this is also accompanied by a breakdown 
of the fluctuation-dissipation theorem, and
the lack of a well-defined free energy.
\bibitem
{intuitive}
For critical properties, this is not a serious problem because
the relevant physics is in the long-wavelength, long-time
limits, where by virtue of universality, 
a kinetic equation postulated solely by 
symmetry considerations is sufficient.  
However, this often requires physical intuition and insight
whereas the present method does not.
\bibitem
{truncate}
Due to factorizations, terms of $O(\phi^n)$ and higher are absent
where $n$ is the number of spins involved in the energetics in $w$
of (\ref{master}). For Glauber dynamics, $n=2d+1$.
\end{references}
\end{document}